\documentclass[conference, onecolumn]{IEEEtran}

\usepackage{graphicx}

\begin{document}

\title{COMODI: Architecture for a Component-Based Scientific Computing System}

\author
{
  \authorblockN{Zsolt I. L\'az\'ar, Lehel Istv\'an Kov\'acs, Bazil P\^{a}rv}
  \authorblockA{Babe\c{s}-Bolyai University, Cluj-Napoca, Rom\^{a}nia\\
  zlazar@phys.ubbcluj.ro, \{klehel,bparv\}@cs.ubbcluj.ro}\\
}

\maketitle

\begin{abstract}
The COmputational MODule Integrator (COMODI) \cite{COMODI} is an initiative aiming at a component based framework, component developer tool and component repository for scientific computing. We  identify the main ingredients to a solution that would be sufficiently appealing to scientists and engineers to consider alternatives to their deeply rooted programming traditions. The overall structure of the complete solution is sketched with special emphasis on the Component Developer Tool standing at the basis of COMODI.
\end{abstract}

\IEEEpeerreviewmaketitle


\section{Introduction}

In the last decades computer science has undergone a spectacular development that has left its marks on most aspects of our everyday life while truly revolutionizing a number of domain specific human activities. However, brute computing force growing at exponential rate was not the answer to the problem of complexity as signaled already in the late sixties when the term ``software crisis" got coined \cite{Naur69}. Thanks to a dynamic IT market marked by fierce competition, software development has been constantly maturing ever since. Unfortunately, computational science remained immune to all those environmental factors that drive the development of the software industry. Even though it has always made much better use of the improving hardware capabilities by squeezing every last drop of performance out of the available computing resources, in terms of efficiency and software quality it has fallen much behind modern trends. 
Recently, almost forty years later, scientists realized too the obstacles standing in the way of large scale scientific software projects \cite{Post04,Post05}. The two papers send a clear message to the computational science community: \emph{a change of paradigm is necessary!} In \cite{Lazar04a} and \cite{Lazar05cse} this message is reinforced in the context of small and medium size projects that make up the bulk of the activity in the community. It is pointed out that there is an acute need for shifting towards a reuse oriented paradigm which would improve dramatically the efficiency and quality of computational research. 

Component software \cite{Szyperski02} is a relatively recent buzzword in the software industry many viewing it as the holy grail of software reuse. These ideas found a fertile soil in computational science thanks to grid computing that enjoys high priority on the research agenda of most developed regions such as EU, North-America and Japan. Grid computing, even though still very much in its infancy, has been preparing the path for component software for some years.

\section{Babel and the Scientific Interface Definition Language (SIDL)}

Driven by the challenges of grid computing the fundamental ideas from component software have quickly been taken up by a few groups and a recommendation for a Common Component Architecture (CCA) emerged \cite{Armstrong99}\cite{CCA_00}. The dominance of this approach is secured by the Common Component Architecture Forum, an organization backed by several US based universities \cite{CCAF}. At the core of CCA stands a language interoperability tool, called Babel, and the Scientific Interface Definition Language (SIDL)\cite{Babel}. Starting from a manually written SIDL file Babel can generate stub and skeleton source code for a number of programming languages commonly used in scientific computing. The programmer can include the implementation part and after compilation the functions can seamlessly communicate across the boundaries of programming languages. Several toy and production frameworks have been developed based on Babel \cite{CCAFE}\cite{XCAT}\cite{SCIRUN2}. They provide a visual programming environment for assembling components into computational projects. There is no doubt that this approach can be viable on the long run yet presently it is not much more than an intellectual experiment as the amount of training for a developer trying to produce a CCA compliant component represents a threshold that are too high for most to consider the solution. The developer is required to learn SIDL and the constituting object-oriented programming concepts. Existing numeric libraries cannot be turned into a component without carrying out changes in the function signatures. Beside the involved extra work, another reason  why many would not take this step is because, as long as CCA is not an established standard, ``spoiling" the code's compliance with previous prescriptions would not pay off. Especially not for a natural scientists or engineers who want the ``most bang for the buck" in their own research and do not share the excitement of computer scientists over using a well designed but fresh standard that apparently makes life only more difficult for quite a few years to come. No wonder that the associated public component repository, Alexandria, is still empty several years after project launch. 

\section{IRIS Explorer}

Another notable solution is the IRIS Explorer of the Numerical Algorithms Group (NAG)\cite{NAG}. Apart from the wide-spread flow-based visual programming environment on the user side, on the component development side the author is assisted by a module builder application that, after collecting sufficient information on the input and output ports via self-explanatory GUI forms, it wraps and builds the code into an IRIS module. However, documenting the code in terms of its interface is completely manual and avoiding mismatches between the source and its description is the responsibility of the author. Using custom datatypes constitutes another challenge for the developer. The two main reasons, though, why the NAG solution fails to have the impact we envision for the future computational framework is that, on the one hand it is limited to mathematical problems and data visualization, on the other it is a commercial product. As a result, in spite of the IRIS Explorer Centre of Excellence initiative \cite{NAGExcellence} the public repository has not received new modules since 2002. Nevertheless, since the NAG solution is endowed with a multitude of deasirable features, we consider that its user base should have an important word to say in conceiving the new high impact solution.

\section{COMODI: Premises and commandments}

Determined to avoid the aforementioned pitfalls the COMODI initiative is meant to be a pragmatic approach with well-defined objectives formulated in view of the ultimate goal of moving computational science out of its nadir and giving it a hand in starting on the road towards modern programming practices while harvesting along the way all the efficiency and quality benefits of a new reuse oriented paradigm. COMODI emerges from a set of premises that are not exclusively its own. Even though not always explicitly formulated by the authors of other solutions yet stand at the basis of these approaches. The following premises are the distilled conclusions of a preliminary survey made with a mixed group of computational scientists on the occasions of conferences, workshops and via an online form on the COMODI website \cite{COMODI}:
\begin{enumerate}
\item computational science is not computer science: the roots of computational science are in natural sciences and engineering not in computer science;
\item computational science demands a new paradigm. The efficiency and quality of the scientific software development process and the reproducibility of virtual computer ``experiments" need major improvements.
\item computational science will not take up any new technologies in the short run unless simpler than those that are in use today;
\item the community itself should decide on the new paradigm. A restricted group of promoters can only set the process off and catalyze it;
\item a few, relatively homogeneous groups make up the bulk of the community. The new solution should target these groups instead of trying to be fully comprehensive.
\end{enumerate}
Out of the above five points the last three are COMODI specific but the extra emphasis on point 3 is the one that makes COMODI essentially different from the rest. This latter postulate translates into a fundamental requirement that we can term as \emph{``zero effort threshold"}. We can make it more specific by formulating a series of commandments that will guide the design process. A clear distinction needs to be made between what COMODI means to end-users on the one hand, and to component developers or authors on the other. The former group is primarily involved in assembling components into projects and executing them while component authors design and implement new components. Since the two activities require different skills and work methods the requirements set for the employed tools in each case also differ. For user satisfaction the solution has to be endowed with the following features:
\begin{itemize}
	\item user friendly graphical interface;
	\item intuitive representation of data and processes such that the elements of low-level programming, C and Fortran programmers  are accustomed to, can be clearly identified;
	\item high-level flow-based visual programming environment;
	\item possibility for low-level control;
	\item platform independence of the framework and of the components;
	\item comprehensive component repository;
	\item FREE!
\end{itemize}
In order to fully support developers it is imperative that no compliance criteria are set for the computational code neither in terms of structure nor used data types. In other words, any valid code written in the supported programming languages should automatically be ready for COMODI. Therefore the following restrictions apply to adapting regular code to COMODI:
\begin{itemize}
	\item no change in the source code. Neither in the interface nor in the implementation;
	\item no extra coding. Connectivity is achieved by supplementing author provided source-code with automatically generated glue-code; 
	\item no need for the author to know other languages/standards then the ones used for implementing the code;
	\item no platform dependence. The capabilities of the system the development is carried out on is extended by on-line servers providing compilation as web service;
	\item no language dependence. C/C++, Fortran, Java, Python and other present and future languages should be able to communicate seamlessly;
	\item low performance overhead; 
	\item support for both open source and commercial components.
\end{itemize}

For closely matching the low-level approach of computational scientist's to programming, components need to be the lowest possible granularity units of code, namely functions and procedures. In \cite{Lazar04a} and \cite{Lazar04b} they are also referred to as \emph{logical components} so that they could be clearly distinguished from higher granularity \emph{physical components} or \emph{packages} that are units of deployment. A physical component can contain a large number of logical components that are packed together and then uploaded as one file into the repository. 

Clearly, the above commandments are easier to state than to comply with. Communicating data across, language and platform boundaries represents a serious technical challenge. The tip of the iceberg includes the following issues:
\begin{itemize}
	\item language dependence: e.g. Pascal uses the call stack differently; 
	\item architecture dependence: file systems, little endian vs big endian; 
	\item compiler dependence: the same data type may be represented differently. e.g. 16 bits vs 32 bits;
	\item exporting type definitions.
\end{itemize}
For bridging over languages there are several solutions at hand such as translating all types of source code into a single language, e.g. C, or compile them to an intermediate language similar to Java bytecodes or the Microsoft IL for .NET. Alternatively, following the design of IRIS Explorer one can provide an API for the different languages or create stub and skeleton code based on some interface description as done with Babel for the Common Component Architecture (CCA). However, none of the above can fully live up to the expectations formulated in the commandments from the previous section as they either require extra care from the developer or interfere with his/her source code or induce a significant performance overhead. We suggest that the responsibility of all wiring related issues should be assumed by generated glue-code. Similarly to IRIS and unlike Babel, interface glue-code is generated after the implementation which apparently is not a healthy programming practice as this would prescribe defining all interfaces first and then stuffing the implementation into the code. However, Babel glue-code is generated from an SIDL file written manually by the developer. This means a new language to be mastered which would make COMODI much less appealing. Besides, the target segment of COMODI are scientists that already have programmed for a few years and ready to try COMODI in view of the zero effort threshold. These either already have their implementation ready or will develop it using their own familiar programming environment. IRIS does a better job in this respect but falls short of effectively minimizing the effort of the developer and when it comes to user defined datatypes the developer bumpes into APIs and an involved type definition procedure.

Thus, the problem of interoperability in itself is not impossible to circumvent. The question remains though -- at what cost? By ``cost" we mean the compromises that need to be made when there are no available alternatives for simultaneously respecting all of the above formulated general requirements to their full extent. Our small prototypes arm us with confidence but complete certainty is conditioned by a solution that has been tested against a wide range of different computational tasks. 

Merging the reuse oriented paradigm with that of distributed computing represents a double effort threshold for the community, on the user and developer side alike. Both areas are in an experimental stage. Therefore we recommend that COMODI should focus on providing a viable solution for the problems it is targeting, namely efficiency and quality of scientific software development, without getting involved in the problem of efficient computing and storage resource sharing. Nevertheless, all design decisions should be taken in view of the requirements for adapting the framework to the grid in the near future.

\section{Architecture}

The complete solution is made up of the following major elements:
\begin{itemize}
\item high-level visual programming environment for computational projects;
\item standardized scientific component descriptor language (CDL);
\item component developer tools for adapting regular code to the framework;
\item distributed component repository;
\item compilation web service.
\end{itemize}

\begin{figure}[htbp]
\centerline{\includegraphics[width=12cm]{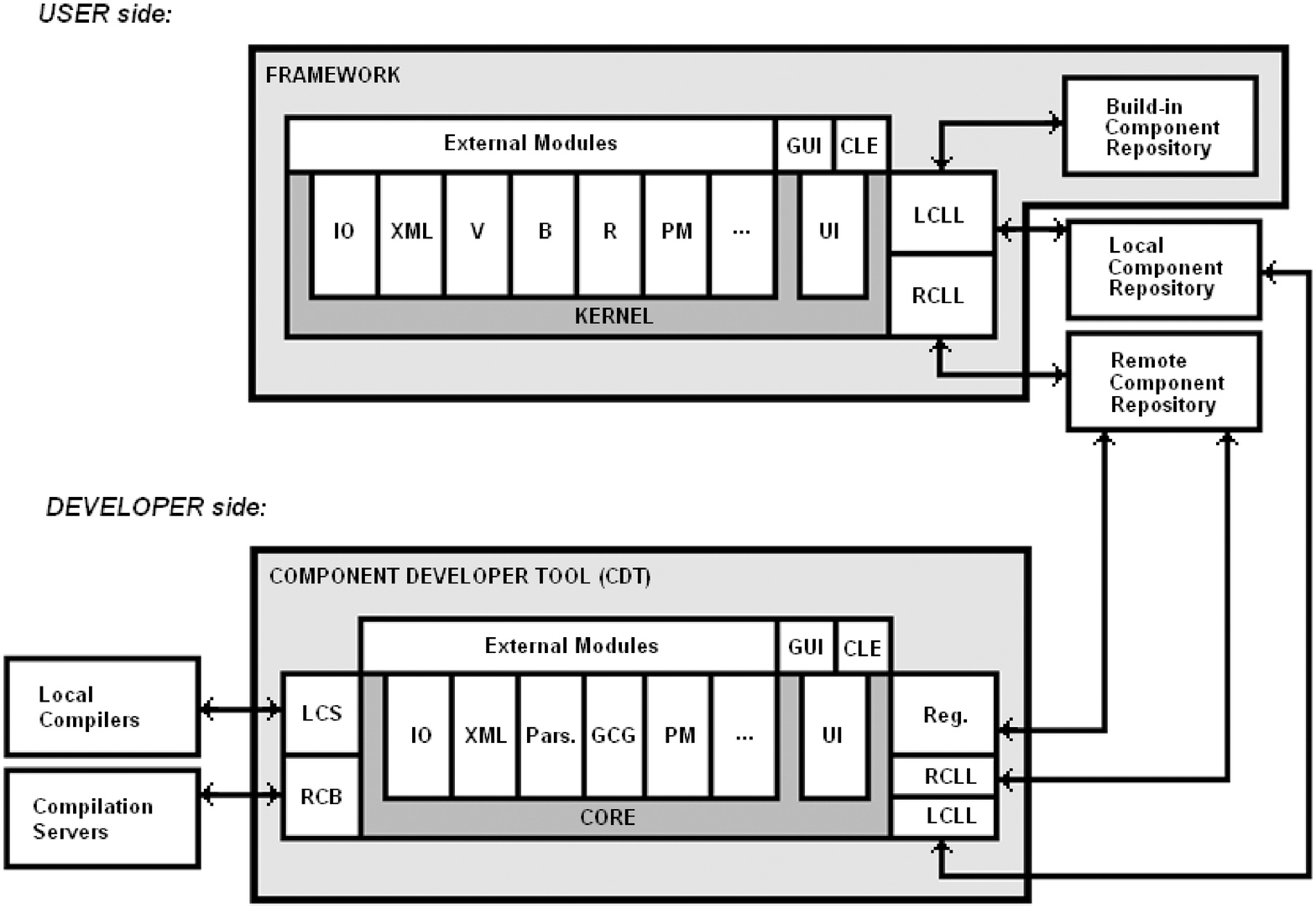}}
\caption{On the user side: IO: \textit{Input/Output System}, XML: \textit{Extended Markup Language Parser}, V: \textit{Validator}, B: \textit{Binding System}, R: \textit{Running System}, PM: \textit{Project Manager}, UI: \textit{User Interface}, GUI: \textit{Graphical User Interface}, CLE: \textit{Command Line Editor}, LCLL: \textit{Local Component Locater and Loader}, RCLL: \textit{Remote Component Locater and Loader}.
On the developer side: LCS: \textit{Local Compilation Service}, RCB: \textit{Remote Compilation Broker}, Pars.: \textit{Parser}, GCG: \textit{Glue-Code Generator}, Reg.: \textit{Registrar}.
}
\label{F:architecture}
\end{figure}

As it has already been suggested in the previous section, depending on the programming activities the above software elements support they can be divided into two fundamental groups: the \textit{user side} and the \textit{developer side}. On figure \ref{F:architecture} we can see the sketch of the COMODI architecture. The responsibilities of each part are summarized in Table \ref{T:roles}.
\begin{table}[!htb]
\begin{center}
\begin{tabular}{l|l}
	{\bf Role of the framework} & {\bf Role of the component developer tool} \\
	\hline
	\parbox[t]{6.5cm}{
	\begin{itemize}
		\item component assembling
		\item project verification and validation
		\item project execution
		\item runtime user interaction
	\end{itemize}
	}
	&
	\parbox[t]{6.5cm}{
	\begin{itemize}
		\item assist the developer in documenting the code
		\item generate glue-code
		\item assist the developer in compiling the component
		\item register the component with the global repository
	\end{itemize}
	}
	\\
\end{tabular}
\end{center}
\caption{Responsibilities of the two major parts of COMODI}
\label{T:roles}
\end{table}
The developer layer contains a user friendly \textit{Graphical User Interface} (GUI), a \textit{Component Developer Tool} (CDT) with a \textit{Parser}. The CDT, after semi-automatically collecting information pertaining to the content, behavior, and representation of the component, generates a \textit{component descriptor file} (CDF) in the XML based \textit{Component Desciptor Language} (CDL) and the source of the \textit{glue-code} that will intermediate the communication of the component within the COMODI framework. At this stage the CDF will contain all communication related information such as exported functions and data types. It describes both syntactically and semantically the component, supports the programming style of computational scientists as far as data structures, and it is extensible. Its complexity is expected to grow together with the user community and the number of application areas. By semi-automatical we mean that the \textit{Parser}, which stands at the basis of the tool, inspects the source file and generates a primary CDF. Using the  GUI, the developer only has to confirm the exported ports, provide human readable documentation for the component, set default values and add representation related information. The CDT then contacts on-line \textit{compilation servers} and returns ready-made binaries for the platforms of the developer's choice. The compiled library together with the descriptor file is uploaded by the developer to a place where it can be accessed publicly while the CDT registers the component in the \textit{component repository}. 

The deployed component is a package containing the component's source code - if the developer chooses to make the source open - the component descriptor file, the binaries for both the computational- and the generated glue-code, and further resources. Components are packed into standard ZIP or TAR.GZ format and registered in the \emph{Remote Component Repository}. Upon use within the COMODI framework, the component is downloaded and stored in the \emph{Local Component Repository}.

The sources provided by the component developer suffer no changes whatsoever during the component creation process. All glue-code comes as additional functions in a separate file. Not touching the source of the developer has the benefit of the compiled component being usable both within and outside the COMODI framework making COMODI components fully compatible with traditional programming environments.


In order to make COMODI itself easily extensible it has to be component-based. This requires the separation of the framework into a \textit{kernel layer} and several other modules built on the top of it. It is possible to enforce a very general view on this component architecture and deal uniformly with computational components and components that are intimately related to the framework itself. In this approach, anything apart from the kernel is a component, be it a simple numerical component or a heavyweight GUI. However, this uniformity, while simplifying the integration of components vital to the proper functioning of COMODI, will come dear as it compromises the postulated simplicity of wiring computational components by users. Therefore it is sensible not to sacrifice the support of user and component developer activities in favor of those related to the development of COMODI \cite{Lazar04a}.

\section{Glue-code and connector components}
In the previous section we pointed out that ``clever'' glue-code is the key to following the COMODI commandments. It can come in two flavors depending on whether it mediates an incoming function call through a provides port or an outgoing call via a uses port. We shall prepend the term glue-code with the words "uses" or "provides" whenever this aspect will be of relevance. Connectors represent a similar concept intimately related to component software. There is no fundamental difference between connectors and glue-code. Both are meant to bridge over incompatibilities that are not essential from the point of view of the composed client-server system. As such they can be handled automatically or semi-automatically. The most important difference between the two consists in the fact that glue-code is generated while connectors, as components in general, are hand-made. Glue-code is tightly associated with a component within the boundaries of the same physical component (deployed unit). Glue-code is a kind of integrated connector that can communicate with anybody at one end but it only connects to a well-defined component located in the same physical component. 

Glue-code is included into the component instead being part of the framework for a number of reasons:
\begin{itemize}
	\item there are functionalities that can only be set statically, depending on the content of the source code;
	\item it can be better optimized for performance: in order to satisfy the requirement of low overhead it is necessary that the framework does not intermediate the communication between components. Instead, it will wire up the connections by setting direct component-to-component references \cite{Lazar04b}; 
	\item freedom of the component author to further optimize it;
	\item keeps the framework platform independent;
	\item in view of the approaching era of grid computing the autonomy of components should be increased.
\end{itemize}
The obvious disadvantage of this solution is the components' increased size. However, this increase is not expected to be a relevant problem.

Since glue-code takes over all the burden of making regular code connectible it has several responsibilities: 
\begin{itemize}
	\item {\em call stack management:} bridge the difference in handling the call stack in various languages
	\item {\em parameter passing management:} when the two involved languages can't automatically do it by themselves. For example, when calling Fortran from C, parameters can only be passed by reference.
	\item {\em linking:} by linking we mean the process of setting all call references between components. These references are set according to parameter strings extracted from an XML encoded file containing the description of the assembled computational application. They are stored in static variables of glue-code segments and are passed to the computational code during runtime (see figure \ref{F:linking}). The runtime entry point is called once also during link-time and the children references are requested from the glue-code function in charge with wiring. As a result, the wiring function can be completely avoided during runtime. The actual computational code, referred to as ``business logic" in figure \ref{F:linking}, will be in the body of a function that receives all necessary information, including the child references, as parameters. This has the additional benefit of self-containment. The function is also fully functional  outside the context of the framework  without confusing extra arguments. A more detailed discussion on the wiring mechanism can be found in \cite{Lazar04b}.
\begin{figure}[htbp]
\centerline{\includegraphics[width=6cm]{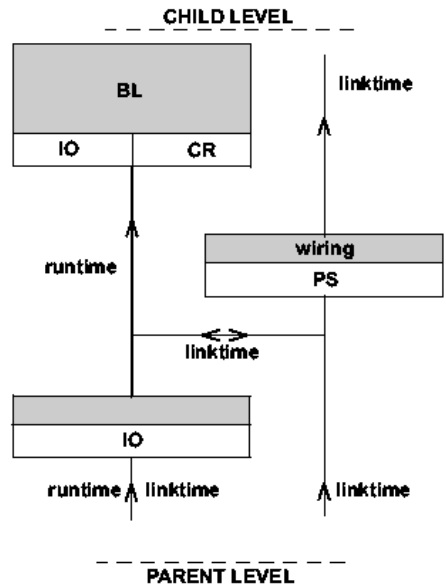}}
\caption{Glue-code aided linking of components. Notation: PS = paramString, IO = input/output, BL = business logic, CR = children references}
\label{F:linking}
\end{figure} 	
	\item {\em handling data and code aggregates:} translates between data structures of different granularity. For example, structures  records and objects can be decomposed into a set of variables and vice versa. This is especially important when binding C++ code to C.
	\item {\em managing default values and references:} the author of a component is encouraged to provide default values for the parameters of both uses and provide ports. There are at least two benefits to this feature. For once, users can gradually explore the capabilities of a component by using a reduced set of parameters at each port. Secondly, default parameters endow a given port with the flexibility of connecting to ports that require a different number of parameters.
	\item {\em managing global naming conventions:}
	since the components are autonomous, authors must respect only the rules of the programming language in which the component is developed. Problems may arise with the identifiers (names of functions, procedures, types, variables, constants, parameters) -- if two or more authors give the same name to two or more entities. The glue-code generator translates the identifiers between local and global naming conventions.
	\item {\em managing remote calls:} it realizes a \emph{stub-skeleton} architecture for local and remote component calls. 
\end{itemize}
The details implementing connectors in COMODI is a topic on its own right and it is beyond the scope of this paper. The reader can refer to chapter 10. in \cite{Szyperski02} and \cite{Mayer02}.

\section{The parser}

One of the particularities of the COMODI solution is that the process of converting the source code into a component is automated to the maximum possible extent. Therefore the component developer tool must ``gain insight'' into the the source code with the help of the parser. The aim of the \textit{parser} is to extract the necessary information from the programs source code in order to elaborate a full documentation of the component. The output of the parser is an XML document that is piped into the glue code generator and also serves as raw material for the content of the component descriptor file. The parser analyzes lexically and syntactically the source of the developers computational code. The semantic analysis is outside the scope of the parser. 

The used information is:

\begin{itemize}
	\item identifiers
	\item types
	\item variables
	\item constants
	\item functions
	\item procedures
	\item parameters
	\item comments for the documentation
	\item special directives
\end{itemize}

\begin{figure}[htbp]
\centerline{\includegraphics[width=12cm]{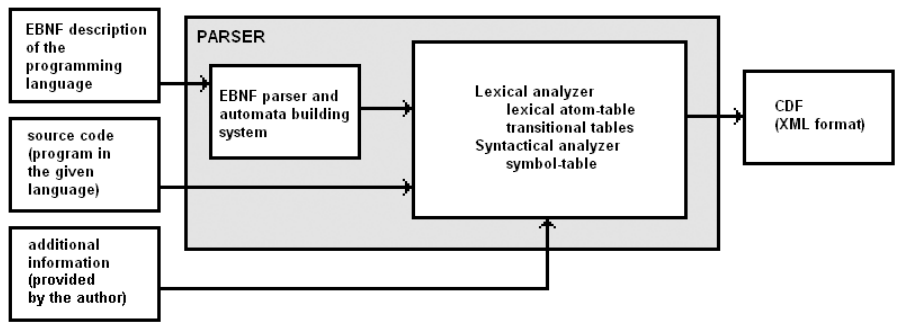}}
\caption{Structure of the parser}
\label{F:parser}
\end{figure}

The architecture of the COMODI parser consists of two blocks: \textit{the programming language description and recognition part} (EBNF parser) and \textit{the program recognition part}, which carries out the lexical and the syntactical analysis based on a self-constructive automata-system (Fig. \ref{F:parser}).

The description of the programming languages is given in EBNF format (Standard: ISO/IEC 14977).
The \textit{EBNF parser} reads the EBNF description of the programming language (Fortran, C, 
C++, Pascal, Java, etc.), and builds an automata-system, responsible for the lexical and the syntactical analysis of the source code.

The automata-system contains a set of modified push-down automata, interconnected into a network, each automaton each possessing an inner stack. This system is equivalent to a single non-deterministic push-down automaton, capable of recognizing context-free languages. The lexical elements and the syntax of the majority of the used imperative programming languages, and also their recognition rules, can be defined with context-free grammars. So, our automata-system can recognize the majority of imperative programming language, once their EBNF definitions are available.

The automata-system receives as input the program source code -- each token is an element of the alphabet -- starts from an initial state, and after reading the symbol next in the line, it changes the inner-state according to the transitional function. During this transition  it modifies the stack. After reading all the input symbols, if the automata-system is in one of the final states, and the stack is empty, the automata-system recognizes the input source code and builds correctly the tables. In all 
other cases it fails.

Our prototype show that the above algorithm seems to work well for languages such as C or Fortran. 

\section{Conclusions and outlook}

We have presented the general requirements and a few design guidelines for a complete reuse oriented solution for computational scientists. The lack of impact of present solutions is blamed on the involved high effort threshold, in many cases made worse by a closed source and restrictive copyrights.  Therefore, the corner stone of COMODI is the zero effort threshold requirement for component developers. We argue that the community needs a solution that allows a smooth, effortless transition to the new paradigm. Once there, high-tech solutions will automatically be accepted by the community. We also claim that a change of paradigm requires a solution that is widely used and supported, situation that is only conceivable within an OpenSource project. COMODI should be a joint effort of computer scientists assuring the quality of the code and computational scientists collectively and actively contributing to refining the requirements. If the contribution of either of the two sides gets out of balance COMODI ends up as just another interesting case study in computer science or, conversely, becomes an unreliable pile of code impossible to maintain. Our prototypes indicate that the suggested architecture is feasible. Thus, the main challenge is not of technical nature but rather consists in sparking the interest of the community in developing and using COMODI.


\section*{Acknowledgment}
This work is supported by the National University Research Council of Romania with grant no. 27687/14.03.2005.






\end{document}